\documentstyle[psfig,12pt,twoside]{article}

\begin{document}

\date{}
\baselineskip=0.76truecm
\evensidemargin=1.2truecm
\oddsidemargin=1.2truecm
\topmargin=0.8truecm
\textheight=20.0truecm \textwidth=14.0truecm

\titlepage
\title{Hadronization mechanisms and\\
longitudinal polarization of 
$\Lambda$ in $e^+e^-$ annihilation at high energies
\footnote{Work supported in part by 
the National Natural Science Foundation of China} }
\author {Liang Zuo-tang \\
    Department of Physics,
   Shandong University, \\ 
  Jinan, Shandong 250100, CHINA}

\maketitle

\begin{abstract}                

We suggested that longitudinal polarization 
of Lambda produced in 
$e^+e^-$ annihilation at LEP energies  
can provide useful information on 
hadronization mechanism in general and 
on testing the validity of 
different pictures for the spin content 
of baryon in describing 
the fragmentation process in particular.
We present the results 
obtained from the calculations 
based on two very much different pictures.
We compare the results with the recent ALEPH data    
and make suggestions for future measurements.

\end{abstract}     

\newpage
In this talk, I would like to discuss a question which we 
met in our study on spin effects in hadron production processes 
and the efforts we made toward solving this problem.
The work which I will present in the following has been done 
in collaboration with Dr. Boros. The main ideas and results 
have already been published in [1].

As you certainly know, 
there exist now in literature two completely different 
pictures for the spin contents of the baryons:
According to the static (or constituent) quark model, 
spin of a baryon belonging to the $J^P={1\over 2}^+$ 
octet is completely determined by 
the three valence quarks. 
The sum of the spins of these valence quarks 
is the spin of the baryon. 
This picture is very successful in describing 
the static properties of the baryons.
But according to the 
the data from polarized deep inelastic 
lepton-nucleon scattering [2]
and SU(3) flavor symmetry in hyperon decay, 
the sum of the spins of the 
three valence quarks is only 
a small fraction of the 
spin of the nucleon.  
A large part of the baryon spin originates from the 
orbital angular momenta of the valence quarks and/or 
from the sea (i.e. the sea quarks, antiquarks and gluons).
Hence, we met a question:
Which picture is suitable in describing 
the spin effects in the fragmentation process?   
Obviously, the answer to 
this question can be different in different 
hadronization models. Hence, study 
of this question should be able to 
provide also further test of these models.
Usually, we say that, in the hadronization process, 
a quark and an antiquark combine to form a meson 
or three quarks combine together to form a baryon. 
In this way, we consider only the valence quarks of 
the hadron but neglect the sea. 
It seems that here the SU(6) wave function 
based on the static quark model should 
be suitable. 
Surely, whether this is indeed true 
is a priori unknown
and should be studied in experiments. 
We found that 
polarization of Lambda is an ideal place to 
investigate this problem because of the following:  
First,  the spin structure of Lambda 
in the static quark model is very special:  
the spin of Lambda is completely carried by 
the $s$ valence quark 
while the $u$ and $d$ have completely no contribution.
This picture is completely different from that 
drawn from the data of 
deep-inelastic lepton-nucleon scattering [2] 
and SU(3) flavor symmetry in hyperon decay.   
The deep inelastic scattering data, 
together with the SU(3) flavor symmetry for hyperon decay,
suggest that [3] the $s$ quark 
carries only about $60\%$ of the Lambda spin, 
while the $u$ or $d$ quark each 
carries about $-20\%$. 
Second, the polarization of the produced Lambda 
can easily be determined 
in experiments by measuring the angular distribution 
of the decay products. 
Besides, striking polarization effects have been observed 
for hyperons  produced in unpolarized 
hadron-hadron collisions experiments [4].
Such effects have been observed  
for more than two decades and remain as 
a puzzle for the theoretians. 
Clearly, the study of the above mentioned 
question should be able to provide some useful 
information of this problem; and this 
makes the study even more 
interesting and instructive. 

Polarization effects for Lambda 
produced in high energy reactions
have been studied in different connections [3,5-13].
In some of these discussions [3,5-10], 
current quark picture has been used 
thus the picture for the spin content of Lambda 
drawn from the polarized deep 
inelastic lepton nucleon scattering data 
should be applicable.
But in the other [11-13], 
it is assumed that Lambda spin is completely determined by the 
$s$ quark thus picture of the static quark model 
should be applicable. 
No discussion has been made yet to the question 
of which of them is more suitable.

It is known from the standard model of 
electroweak interaction that 
the $s$ quark produced in $e^+e^-$ annihilation 
at high energies is longitudinally polarized [14]. 
Hence it is expected [14] 
that the Lambda which contains this $s$ quark 
should also be longitudinally polarized 
and such Lambda polarization can be measured in experiments. 
Theoretically, this Lambda polarization can be calculated 
and the results should be quite different using 
the above mentioned two different pictures 
for the spin contents of Lambda. 
Hence, measurements of the polarization should be able to 
show which picture is more suitable in describing such spin effects.
Calculations of the longitudinal 
Lambda polarization in $e^+e^-$ 
annihilation at the $Z$-pole 
has been made [15,1] using the picture of 
the static quark model, and 
using the picture drawn from the data of 
deep inelastic scattering respectively.
These calculations are in principle exactly the same. 
I will therefore briefly summarize  
the calculations made in [1] 
and their cmparison with those in [15] 
and the available data [16] in the following.

Here, we first consider the contribution of the Lambdas 
which are directly produced in the hadronization process.
Such hyperons are divided into two groups: 
those which contain the leading $u$, $d$ or $s$ quark 
and those which do not. 
The former kind of Lambdas, i.e. those which contain 
the initial $u$, $d$ or $s$ quark from $e^+e^-$ annihilation, 
can be polarized since 
the initial $u$, $d$ or $s$ quark is 
longitudinally polarized.
But the latter are assumed not to be polarized. 
This is not only true in the popular hadronization
models such as LUND model [17] but also 
consistent with the experimental observations 
that both hyperon polarization in unpolarized 
              hadron-hadron collisions and left-right asymmetries 
              in inclusive production processes in 
              single spin hadron-hadron collisions in the 
              central rapidity region are consistent with zero 
              although they are quite large in the fragmentation 
              region. (See e.g. [4] and the references given there). 
The polarization of former kind of Lambda is different  
in different pictures for the spin structure of Lambda. 
More precisely, the polarization of such Lambda 
is equal to the fraction of spin 
carried by the quark which has the flavor of the 
initial quark multiplied by  
the polarization of this initial quark. 
The polarizations of the initial quarks from 
$e^+e^-$ annihilations are determined 
by the standard model for electroweak interactions, 
and given by [14],
\begin{equation} 
P_f=-\frac{A_f(1+\cos^2\theta)+B_f\cos\theta}
          {C_f(1+\cos^2\theta)+D_f\cos\theta}
\end{equation}
where $\theta$ is the angle between the outgoing quark and 
the incoming electron, the subscript $f$ denotes the flavor 
of the quark and
\begin{equation}
A_f=2a_fb_f(a^2+b^2)-2(1-{m_z^2 \over s})Q_fab_f,
\end{equation}
\begin{equation}
B_f=4ab(a^2_f+b^2_f)-2(1-{m_z^2 \over s})Q_fa_fb,
\end{equation}
\begin{equation}
C_f=\frac{(s-m_Z^2)^2+m_Z^2\Gamma^2_Z}{s^2}Q^2_f+
(a^2+b^2)(a^2_f+b^2_f)-2(1-{m_z^2 \over s})Q_faa_f,
\end{equation}
\begin{equation}
D_f=8aba_fb_f-4(1-{m_z^2 \over s})Q_fbb_f,
\end{equation}
where $m_Z$ and $\Gamma_Z$ are the mass and decay width of $Z$; 
\begin{equation}
a=\frac{-1+4\sin ^2\theta_W}{2\sin 2\theta_W}
\end{equation}
\begin{equation}
b=-\frac{1}{2\sin 2\theta_W}
\end{equation}
\begin{equation}
a_f=
 \left\{ \begin{array}{r@{\quad\quad}l} 
\frac{1-8\sin ^2\theta_W/3}{2\sin 2\theta_W}, & 
 \mbox {\rm for } f=u,c,t,\\ 
\frac{-1+4\sin ^2\theta_W/3}{2\sin 2\theta_W}, &
 \mbox {\rm for } f=d,s,b, 
\end{array} \right.
\end{equation}
\begin{equation}
b_f=
 \left\{ \begin{array}{r@{\quad\quad}l} 
\frac{1}{2\sin 2\theta_W}, & 
 \mbox {\rm for } f=u,c,t,\\ 
-\frac{1}{2\sin 2\theta_W}, &
 \mbox {\rm for } f=d,s,b, 
\end{array} \right.
\end{equation}
are the axial and vector coupling constants 
of electron and quark to $Z$ boson,  
which are functions of the Weinberg angle $\theta_W$. 
(See also table 1 in [14]). 
Averaging over $\theta$, we obtain 
$P_f=-0.67$ for $f=u,c,t$ and $P_f=-0.94$  for $f=d,s,b$. 

The fractional contributions 
($\Delta U_\Lambda$, 
$\Delta D_\Lambda$, and 
$\Delta S_\Lambda$) 
of different flavors ($u$, $d$ and $s$) 
to Lambda spin are calculated using 
the deep inelastic lepton-nucleon scattering data 
on $\Gamma_1\equiv \int^1_0g_1(x) dx$ 
[where $g_1(x)$ is the spin-dependent structure]
and those for the constants $F$ and $D$ in hyperon decay. 
The detailed procedure of extracting the 
$\Delta U_\Lambda$, 
$\Delta D_\Lambda$, and 
$\Delta S_\Lambda$ from the data for $\Gamma_1^p$ for proton, 
and those for $F$ and $D$ is standard and is 
summarized in the Appendix of [1].
The obtained results are given in Table 1. 

{\small
\begin{table}
\caption{Fractional contributions $\Delta U$, $\Delta D$ and $\Delta S$ 
of the light flavors to the spin of baryons in the $J^P={1\over 2}^+$ 
octet calculated  
using the static quark model (static QM) and those obtained using 
the data for deep inelastic lepton-nucleon scattering and those for 
hyperon decay under the assumption that SU(3) flavor symmetry is valid.  
The first column shows the obtained expressions in terms of 
$\Sigma$, $F$ and $D$. 
The static QM results are obtained by inserting 
$\Sigma=1, F=2/3$ and $D=1$ into these expressions 
and those in the third column are obtained by inserting 
$\Sigma =0.28$, obtained from deep inelastic lepton-nucleon 
scattering experiments [2], 
and $F+D=g_A/g_V=1.2573, F/D=0.575$  
obtained [18,19] from the hyperon decay experiments. }
\begin{tabular}{l||c|c|c||c|c|c}
\hline 
 &\multicolumn{3}{c||}{$\Lambda$} &\multicolumn{3}{c}{$\Sigma^0$}\\ \hline 
& & static QM & DIS data & & static  QM & DIS data \\ \hline
$\Delta U$ & $\frac{1}{3} (\Sigma-D)$  & 0    & -0.17 & 
             $\frac{1}{3} (\Sigma+D)$  & 2/3  & 0.36  \\ \hline  
$\Delta D$ & $\frac{1}{3} (\Sigma-D)$  & 0    & -0.17 & 
             $\frac{1}{3} (\Sigma+D)$  & 2/3  & 0.36  \\ \hline
$\Delta S$ & $\frac{1}{3} (\Sigma+2D)$ & 1    & 0.62  & 
             $\frac{1}{3} (\Sigma-2D)$ & -1/3 & -0.44 \\ \hline \hline 
 &\multicolumn{3}{c||}{$\Xi^0$} &\multicolumn{3}{c}{$\Xi^-$}\\ \hline 
&  & static QM & DIS data & & static  QM & DIS data \\ \hline 
$\Delta U$ & $\frac{1}{3} (\Sigma-2D)$  & -1/3 & -0.44 & 
             $\frac{1}{3} (\Sigma+D)-F$ & 0    & -0.10 \\ \hline  
$\Delta D$ & $\frac{1}{3} (\Sigma+D)-F$ & 0    & -0.10 & 
             $\frac{1}{3} (\Sigma-2D)$  & -1/3 & -0.44 \\ \hline
$\Delta S$ & $\frac{1}{3} (\Sigma+D)+F$ & 4/3  &  0.82 & 
             $\frac{1}{3} (\Sigma+D)+F$ & 4/3  &  0.82 \\ \hline 
\end{tabular}
\end{table}
}

We next consider the contribution 
of those Lambda's from 
the decay of other hyperons in the same octet as Lambda. 
These hyperons can also be polarized if they contain the 
initial $u$, $d$ or $s$ quark, and the polarization 
can be transferred to Lambda's in the decay processes.
The polarization of such Lambda is thus equal to the 
polarization of the hyperon multiplied by the probability for 
the polarization to be transferred to Lambda. 
Hence, to calculate such contribution, 
we need to calculate the polarization 
of the such hyperon before it decays and 
the probability for the polarization 
to be transferred to Lambda in the decay process. 
The polarization of hyperon in the same octet as Lambda 
can easily be calculated using exactly the same method 
as that for Lambda. 
There are three such hyperons, i.e. 
$\Sigma^0$, $\Xi^0$ and $\Xi^-$ which may decay into $\Lambda$.
We calculated the fractional contributions 
of different flavors of quarks to their spins 
in the same way as that for Lambda and obtained 
the results shown in Table 1.
These results are as reliable as those for Lambda, 
and are therefore [3] as reliable as those for the nucleons.
$\Sigma^0$ decay into $\Lambda$ by emitting a photon, i.e.,  
$\Sigma^0\to \Lambda \gamma$. 
Whether the polarization of $\Sigma^0$ is 
transferred to the produced Lambda in this 
decay process has been discussed in [20]. 
It has been shown that, on the average, 
the produced $\Lambda$ is also polarized 
(in the opposite direction as $\Sigma^0$)
if $\Sigma^0$ was polarized before its decay,  
and the polarization is $-1/3$ of that of the $\Sigma^0$. 
The hyperon $\Xi$ decays into $\Lambda$ through 
$\Xi \to \Lambda \pi$, which is a parity 
non-conserving decay and is dominated by S-wave. 
The polarization of the produced $\Lambda$ is equal to 
that of the $\Xi$ multiplied by a factor $(1+2\gamma)/3$, 
where $\gamma$ can be found in review of particle properties [18] 
as $\gamma = 0.87$.

Unfortunately, at present, 
it is only possible to estimate the 
contribution from the decays of 
hyperons belongs to the baryon decuplet
using the picture of the static quark model.
It is however impossible to calculate 
the polarization of the produced 
decuplet hyperons in a way consistent 
with that for those in the octet using 
the picture derived from 
deep-inelastic scattering data [2].
This is because no deep-inelastic 
scattering data on any one of such 
decuplet baryons is available.
It is therefore impossible to calculate 
the fractional contributions 
of different flavors to the spin of such hyperon. 
Hence, it is impossible to estimate the
contributions of decays of such hyperons 
which contain the initial $u$, $d$ or $s$ quark 
to the polarization of Lambda in 
the final state of $e^+e^-$ annihilations 
in the same way as that for the octet hyperons. 
Qualitative analysis suggests that
the influences of such hyperons 
should not be very large.
This is because, first, their production rates 
are relatively small, and  
second, since the mass differences 
between such hyperons and Lambda are relative large, 
their decays contribute mainly to 
Lambda's in the central region of the $e^+e^-$ annihilation 
(i.e. those with relatively small momenta).  
This region is dominated by those Lambda's 
which do not contain the initial quark 
and are unpolarized.

To make a quantitative estimation, 
we need a hadronization model 
to calculate all the different contributions to the Lambda's 
from all the different sources discussed above. 
For this purpose, we used 
the LUND model [17] as implemented by JETSET [21]. 
We explicitly calculated the different contributions, 
and obtained the results shown in Fig.1. 
We see in particular that the contribution 
from the decay of the decuplet hyperons 
is indeed relatively small. 
We calculated Lambda polarization $P_\Lambda$ 
for the following two cases: 
In the first case, we completely neglect  
the contribution from decuplet hyperon decay to $P_\Lambda$  
and obtained the results shown by the 
solid line in Fig.2. 
In the second case, we used the results 
for the polarization of the decuplet hyperons 
obtained from the static quark model 
as an approximation to estimate the contribution 
of such hyperon decay to $P_\Lambda$. 
We added the results to $P_\Lambda$ and obtained 
the dashed line in Fig.2. 
For comparison, we included  in the figure 
also the results from the static quark model 
without (dotted line) or with (dash-dotted line) 
the contributions from decuplet hyperon decay.

\begin{figure}
\vspace{3.4truecm}
\psfig{file=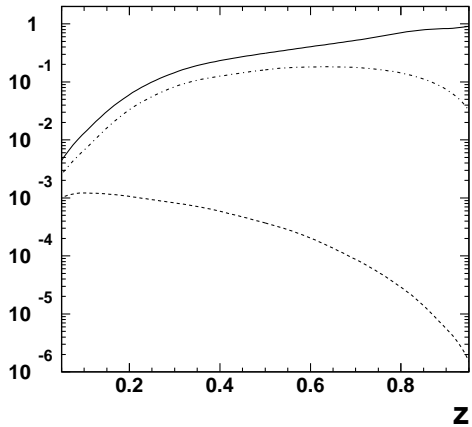,height=5.6cm}
\vspace{-3.0truecm}
\caption{Fractional contributions to Lambdas 
produced in $e^+e^-$ annihilation 
at LEP energy from different sources:  
The solid line denotes those Lambdas which are 
produced directly and contain the initial $u$, $d$ or $s$ quark; 
the dash-dotted and dashed lines are those from decay of octet 
($\Sigma^0$, $\Xi$) and decuplet hyperons ($\Sigma^*$, $\Xi^*$) 
which contain the initial quarks. 
$z\equiv 2p/\sqrt{s}$, where $p$ is the momentum of 
the produced Lambda and $\sqrt{s}$ is the total center of 
mass energy of the $e^+e^-$ system.}
\end{figure}

\begin{figure}
\vspace{3.4truecm}
\psfig{file=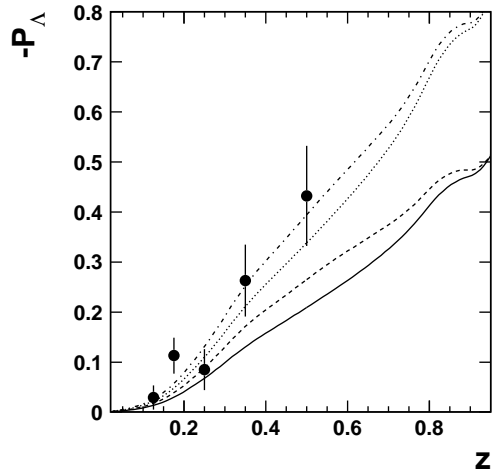,height=7.6cm}
\vspace{-4.60truecm}
\caption{Longitudinal polarization of Lambda, $P_\Lambda$, 
from $e^+e^-$ annihilation at LEP energy as a function of $z$.
(See text for more details).}
\end{figure}

From these results, we see that 
there is indeed a significant difference between 
those obtained in [15] based on 
the picture of the static quark model
and those obtained in the present estimation using a picture 
based on the polarized deep-inelastic 
lepton-nucleon scattering data [2] 
and SU(3) flavor symmetry for hyperon decay.
It seems that the ALEPH data [16] 
favors the former but cannot exclude the latter 
since the error bars are still too large. 
We see also that, although the influence from the decuplet is 
indeed relative small, but it is not negligible 
in particular for moderate $z$. 
We can also see that further measurements of $P_\Lambda$ 
with higher accuracy are needed to distinguish 
between these two kinds of models. 
The large $z$ region is most suitable for such a study  
since in this region not only the magnitude 
of $P_\Lambda$ itself is 
large but also the difference 
between the prediction of the two different models is large. 
It will be also particularly helpful to measure the 
polarization only for those Lambda's 
which are not decay products of decuplet hyperons.

\end{document}